\documentstyle[nato,numreferences,epsf]{crckapb}

\begin{opening}
\title{LINEAR QUANTUM MEASUREMENTS }

\author{D.V. AVERIN}
\institute{Department of Physics and Astronomy, \\ 
Stony Brook University, SUNY \\ 
Stony Brook, New York 11794-3800, U.S.A. }

\end{opening}

\runningtitle{}

\begin{document}

\begin{abstract}

\hspace*{-1.8em} Linear response theory describes 
quantum measurement with an arbitrary detector weakly 
coupled to a measured system. This description produces  
generic quantitative relation characterizing the detector 
that is analogous to the fluctuation-dissipation theorem for 
equilibrium systems. The detector characteristic obtained in 
this way shows how efficient the trade-off is between the 
back-action dephasing and information acquisition by the 
detector. 

\end{abstract} 

\section{Introduction}

If one puts aside philosophical questions created by the perceived 
counterintuitive features of the quantum mechanics that are 
frequently discussed in the literature on the quantum measurement 
problem (an entry point to this literature is provided, e.g., by 
the collections of papers \cite{b1,b2,b3} or monographs 
\cite{b4,b5}) it is easy to see that the physics of quantum 
measurements is fairly well understood by now at least on a 
qualitative level. The process of quantum measurement is dynamic 
interaction between a microscopic quantum system and a 
macroscopic detector that establishes correlations between the 
states of these systems. Although it is impossible to give a 
universal definition of ``macroscopic'' in this context, a 
reasonable working definition is that the macroscopic detector 
is a system with negligible quantum fluctuations. Such systems 
are quite abundant in nature and can be described quantitatively 
within the framework of quantum mechanics. 

Since the description of quantum measurement process as an 
interaction between microscopic and macroscopic systems is quite 
broad, it is of interest to see whether there are any universal 
quantitative features of this process that are independent of  
specific physical realization of the detectors and the measured 
system. The purpose of this work is to show that one such universal 
characteristics of quantum measurements can be obtained from the 
linear response theory. This result was first pointed out in \cite{b6}, 
and this work gives its more concise and accurate presentation.\footnote{ 
{\em Note added in the electronic version.} After this manuscript was 
submitted for publication, the author's attention was drawn to Ref.\ 
\cite{b34} which presents similar development of Ref.\ \cite{b6}. }

An example of the area where a universal description of the 
quantum measurement process can be particularly useful is 
mesoscopic quantum dynamics of solid-state, in particular  
Josephson-junction, qubits. Recent widespread interest in the 
development of solid-state qubits for quantum information processing 
brought with it the discussion of a large number of different 
detectors for measurement of quantum dynamics of these qubits. These 
detectors include quantum point contacts \cite{b20,b21,b22,b23,b24}, 
normal-metal \cite{b25,b26,b27,b28} and superconducting \cite{b29,b30} 
SET transistors operated in different regimes, resonant tunneling 
structures \cite{b31}, generic mesoscopic conductors \cite{b32}, 
SQUIDs \cite{b33}. With such a variety 
of detectors (the list of references above is by no means 
complete), a possibility of giving a quantitative description of 
some detector characteristics independently of its physical 
realization is obviously very helpful for understanding of the 
process of quantum measurement of mesoscopic qubits.

\section{Measurement model and basic relations}

As was discussed in the introduction, the most essential  
feature of the model of the quantum measurement process considered 
in this work is its universality. The model applies to measurement 
with an arbitrary detector which has to satisfy only some basic  
conditions. Schematics of this model is shown in Fig.\ 1 and 
includes the measured system and the detector with the Hamiltonians 
$H_S$ and $H_D$, respectively. The detector and the system 
are coupled through an interaction Hamiltonian $H_{int}$ which 
almost without any loss of generality can be written as the product 
of some system ``coordinate'' $x$ and the detector ``force'' $f$: 
\begin{equation} 
  H_{int} =xf \, .
\label{1} \end{equation}  
The operator $x$ acts as the observable measured by the detector. 
In principle, one could imagine a situation when a detector couples 
to some system through a combination of the several product terms 
like (\ref{1}), but the detector sensitive to several different 
dynamic variables of the system would be quite unusual.

\begin{figure}[htb]
\setlength{\unitlength}{1.0in}
\begin{picture}(5.,.9) 
\put(.75,.05){\epsfxsize=3.8in\epsfbox{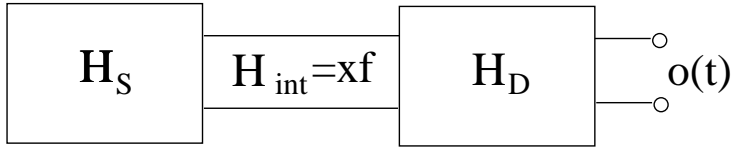}}
\end{picture}
\caption{Schematics of the generic quantum measurement process. 
The detector couples to the measured system via the product of the  
measured observable $x$ and the detector force $f$ and variation 
of its state is reflected in the output variable $o$ with 
essentially classical dynamics. } 
\end{figure}

The general structure of the Hamiltonian of the system: 
\begin{equation} 
H=H_S+H_D+H_{int} \, ,
\label{H} \end{equation}  
is similar to the Hamiltonians studied in the discussions of the 
dissipative quantum dynamics based on the ``system+reservoir'' 
models (see, e.g., \cite{b10}) if the detector is identified with 
the reservoir. This analogy is not accidental, and many features 
of the dynamics of the measurement process are similar to the 
dissipative quantum dynamics of open systems. For instance, as 
will also be discussed below, coupling to the detector leads to 
the dephasing of the measured system, similar to the 
reservoir-induced dephasing.

There are, however, important differencies between the two 
situations. The main difference is that in the case of the 
dissipative quantum dynamics induced by a reservoir, changes in 
the reservoir caused by the system of interest are assumed to be 
unobservable, while the detector as a matter of principle should 
have an output observable $o(t)$ that provides some information 
about its state. To represent the complete measurement process, 
$o(t)$ should behave as a classical variable, i.e. magnitude of 
quantum fluctuations in it should be much smaller than its 
classical component. This condition implies that there is 
another difference between the detector and the dissipative 
reservoir. The detector should be able to convert weak quantum input 
signal $x(t)$ into the classical output signal $o(t)$ that is 
large on the scale of quantum fluctuations. Such a transformation 
requires  amplification and can not be achieved in an equilibrium 
system, i.e., the detector, in contrast to the reservoir has to 
be driven out of the equilibrium. 

In the case of the dissipative reservoirs, it is well known that
the linear response theory is a powerful tool that produces a general 
quantitative statement, fluctuation-dissipation theorem, about the  
reservoir-induced relaxation. The theorem is independent of the 
specific microscopic model of the system. The analogy between the 
measurement and the environment-induced relaxation leads naturally 
to the question of whether some general relations characterizing  
quantum measurements can be derived from the linear response theory. 
As we will see below, the answer to this question is yes, and the 
linear response theory applied to the measurement problem enables 
one to derive generic restriction on the energy sensitivity of a 
general linear quantum detector. The obtained relation characterizes 
how close the detector is to being quantum-limited, i.e., how close 
the detector-induced back-action dephasing in the measured system is 
to the fundamental limit set by the quantum mechanics. While such a 
restriction on the energy sensitivity was known before in several 
specific detector models, linear response theory establishes it for 
an arbitrary detector weakly coupled to the measured system. 

For the linear response theory to be applicable to the measurement 
set-up shown in Fig.\ 1, and to provide some meaningful information, 
the detector has to satisfy certain conditions. Two most important 
are: 
\begin{itemize}
\item the detector is weakly coupled to the measured system so that 
its response to variations in this system is linear,
\item the detector is in the stationary state. 
\end{itemize} 
Since the detector can not be in equilibrium, the second item on 
this list generalizes the equilibrium condition of the standard 
linear-response theory. Both of these conditions are not very 
restrictive, and are satisfied in many experimental situations.

Linearity of the response is not sufficient to specify the detector 
properties completely, since the linear response coefficient can 
have an arbitrary frequency dependence, i.e. one needs to assume 
some spectral characteristics of the detector. Here we consider 
the simplest case when the response time of the detector is much 
shorter that the characteristic time of the dynamics of the measured 
system. This means that the response coefficient $\lambda$ is 
constant in the relevant frequency range. Although this assumption 
is less fundamental that the previous two, and in principle can be 
avoided, it simplifies the discussion considerably. 

Quantitatively, linearity of the detector response means that 
dynamics of the detector operators can be treated in the lowest  
order of the perturbation theory in the detector-system coupling 
$H_{int}$. Expanding the evolution operator $U(t)=e^{-iHt/\hbar}$ 
with the total Hamiltonian (\ref{H}) upto the first order in 
$H_{int}$ we see that the detector output $o(t)$ can be written as: 
\begin{equation} 
o(t)=q(t) + \frac{i}{\hbar} \int^t d\tau [f(\tau),q(t)] x(\tau) \, . 
\label{2} \end{equation}  
The first term $q(t)\equiv e^{iH_D t/\hbar} o(0) e^{-iHt/\hbar}$ 
in this expression can be interpreted as the output noise of the 
detector that exists even in the absence of the input signal, while 
the second term represents the detector's linear response to measured 
system. In zeroth order in coupling, the detector and the system are 
independent, so the averages over their states in eq.\ (\ref{2}) 
can be taken separately. Tracing out the detector variables in 
(\ref{2}) we obtain the regular expression for the linear 
response, extended to the situation when the ``driving force''  
$x(t)$ is a quantum operator. With the adopted assumption of 
instantaneous detector response:   
\begin{equation} 
\frac{i}{\hbar} \langle [f(t),q(t+\tau)] \rangle_D =\lambda 
\delta(\tau -0) \, 
\label{3} \end{equation}  
eq.\ (\ref{2}) reduces to:  
\begin{equation} 
\langle o(t)\rangle_D= \lambda x(t) \, . 
\label{L} \end{equation}  
The infinitesimal shift in the argument of the $\delta$-function 
in (\ref{3}) represents small but finite response time of the detector, 
and is needed to resolve the ambiguity in eq.\ (\ref{2}). In equation 
(\ref{3}), we used the assumption (introduced above) that the 
detector is in the stationary state and the $q$--$f$ correlator 
depends therefore only on the difference of the time arguments, 
and $\langle \dots \rangle_D$ denotes the average of the stationary 
detector density matrix. Since $q(t)$ represents the output noise, 
we took $\langle q(t)\rangle_D =0$. 

Similarly, if the detector produces some non-vanishing average 
force $\langle f(t)\rangle_D \equiv F \neq 0$, it is natural to 
include it in the system Hamiltonian $H_S$, and one can therefore 
always assume that $\langle f(t)\rangle_D=0$. Dynamics of the 
coupling force $f$ generated by the detector is characterized then 
by the correlation function $\langle ff(t)\rangle_D$ and physically 
can be viewed as ``back-action'' noise that dephases the measured 
system. Following the assumption of the instantaneous response 
(\ref{3}) we also take both the output noise and the back-action 
noise to be $\delta$-correlated:  
\begin{equation} 
\langle q(t+\tau)q(t) \rangle_D = 2\pi S_q \delta(\tau)\, , 
\;\;\;  \langle f(t+\tau)f(t) \rangle_D = 2\pi S_f \delta(\tau)\, .  
\label{4} \end{equation}
Here we introduced spectral densities of the output ($S_q$) and 
back-action ($S_f$) noise that are constant in the relevant low 
frequency range. Besides the infinitely small response time of 
the detector, eqs.\ (\ref{4}) also imply that 
dynamics of both $q(t)$ and $f(t)$ is essentially classical, 
since the frequency-dependent part of the spectrum that corresponds 
to zero-point fluctuations gives negligible contribution to the 
correlators (\ref{4}).  

\section{Back-action dephasing versus acquisition of information} 

Qualitatively, measurement dynamics defined by the total 
Hamiltonian (\ref{H}) and assumptions (\ref{3}) and (\ref{4}) 
about the detector characteristics, includes two related 
processes. One is the dephasing and relaxation/excitation of the 
measured system by the back-action force $f(t)$. This part of the 
measurement is very similar to the dissipative quantum dynamics 
of a system weakly coupled to a reservoir. The second process is 
acquisition of information about the state of the measured system 
reflected in the evolution of the detector output $o(t)$. 
On the quantitative level, measurement dynamics depends on the 
Hamiltonian $H_S$ of the measured system and the operator 
structure of the measured observable $x$. 

Probably the simplest situation which can be used to illustrate 
the two qualitative aspects of the measurement dynamics is the 
case of the stationary system, $H_S=0$, with an observable $x$ that has
several discrete eigenstates $|j\rangle$: $x|j\rangle=x_j|j\rangle$  
with eigenvalues $x_j$. Since under the assumption (\ref{4}) the 
back-action force $f(t)$ can be treated as a classical random 
force generated by the detector, one can write the time-evolution 
equation directly for the density matrix $\rho$ of the measured 
system. From the Hamiltonian (\ref{H}), this equation in the basis 
of $x$-eigenstates is: 
\[ \dot{\rho}_{j,j'}=\frac{-i}{\hbar} (x_j-x_{j'})f(t)\rho_{j,j'} 
\, .\] 
Solution of this equation averaged over different realizations 
of $f(t)$ using eq.\ (\ref{4}) gives: 
\begin{equation} 
\rho_{j,j'} (t)=\rho_{j,j'} (0) e^{-\Gamma_d t} \, , \;\;\;\; 
\Gamma_d \equiv  \pi (x_j-x_{j'})^2 S_f/\hbar^2 \, .
\label{5} \end{equation}
From this equation one can see that when the system is prepared in 
the initial state which is a superposition of the eigenstates of $x$, 
i.e. $\rho_{j,j'}$ has non-vanishing off-diagonal elements, interaction 
with the detector dephases the system and suppresses these off-diagonal 
elements. The suppression does not happen only if the states $j$ and 
$j'$ are degenerate with respect to $x$: $x_j= x_{j'}$, so that the 
detector does not distinguish them. Equation (\ref{5}) 
shows also that the rate $\Gamma$ of such a back-action dephasing 
is determined by the spectral density of the back-action force $f(t)$. 

The dephasing (\ref{5}) leads to diagonalization of the density matrix 
in the basis of the measured observable and coincides with the simplest 
instance of the environment-induced dephasing of an open quantum 
system. From the measurement perspective, it describes the dynamic 
side of the ``wave-function collapse'', since when the measurement 
is completed, the measured system should find itself in one of the 
eigenstates of the observable being measured, i.e. $x$. The density 
matrix diagonal in the $x$ representation corresponds precisely to 
such a situation. This interpretation of the back-action dephasing 
is strengthened by the relation between the dephasing rate $\Gamma$ 
and the rate of information acquisition - see, e.g., \cite{b15}. Indeed, 
from this perspective, in the example of the preceding paragraph, the  
detector has to distinguish different eigenstates of the observable 
$x$ and to provide information in which eigenstate the system finds 
itself. According to the linear response 
relation (\ref{L}), the difference between two eigenvalues of 
$x$, $x_j-x_{j'}$, translates into the difference $\delta o= 
\lambda (x_j-x_{j'})$ of the dc values of the detector output $o$. 
The characteristic time $\tau_m$ on which this difference in $o$ 
can be distinguished in the presence of the output noise is 
determined by the low-frequency spectral density $S_q$ of this 
noise. Averaging the $\delta$-correlated noise over the time 
interval $\Delta t$ leaves characteristic noise amplitude 
$\Delta o = (2\pi S_q/\Delta t)^{1/2}$. The characteristic time 
$\tau_m$ of acquisition of the information about the dc value 
of the detector output, and therefore, about the eigenstate 
of the measured observable, is determined by the condition that 
the noise amplitude is reduced at least to half the distance 
$\delta o$ between the signal level. This condition gives 
\begin{equation}  
\tau_m= 8\pi S_q/[\lambda(x_j-x_{j'})]^2\, , 
\label{6}  \end{equation}
and shows that the relation between the back-action dephasing rate 
and the ``measurement time'' $\tau_m$ is independent of the 
eigenvalue difference $x_j-x_{j'}$, and depends only on the 
linear-response parameters of the detector: 
\begin{equation} 
\tau_m \Gamma_d  = 8(\pi/\hbar \lambda)^2 S_q S_f \, . 
\label{7}  \end{equation}
Now the final step is to show that the linear response theory 
relates parameters $\lambda$, $S_q$, $S_f$, in such a way that the 
dephasing rate is fundamentally linked to the measurement time: 
\begin{equation}
\tau_m \Gamma_d \geq 1/2\, .
\label{8}  \end{equation} 

To see this, we start by taking Fourier transform of eq.\ (\ref{3}): 
\begin{equation} 
-i \frac{\hbar \lambda }{2 \pi} = S_{fq}(\omega) - 
S_{fq}^*(-\omega), \, \;\;\; S_{fq}(\omega) =\frac{1}{2\pi} 
\int d\tau e^{i\omega \tau}  \langle fq(\tau) \rangle_D \, .
\label{9}  \end{equation} 
Expanding expression for the correlator $S_{fq}(\omega)$ in the 
basis of the energy eigenstates $|\varepsilon \rangle$ of the 
detector we obtain: 
\begin{equation}  
S_{fq}(\omega) = \int d\varepsilon \rho_D (\varepsilon) 
\nu (\varepsilon)  \nu (\varepsilon- \hbar \omega ) 
\langle \varepsilon|f| \varepsilon - \hbar \omega\rangle \langle 
\varepsilon -\hbar \omega |q|\varepsilon \rangle \, ,
\label{11}  \end{equation} 
where $\nu (\varepsilon)$ is the density of the detector energy 
states, and since the detector was assumed to be in the stationary 
state, its density matrix $\rho_D (\varepsilon)$ in the energy 
basis is diagonal. 

Making use of the expression (\ref{11}) and similar expressions 
for noise spectral densities, we can establish an inequality 
relating them. To obtain it, we note that eq.\ (\ref{11}) can be 
viewed from a somewhat artificial, but useful perspective as a 
scalar product of two functions of energy $\varepsilon$, 
$\langle \varepsilon -\hbar \omega 
|q|\varepsilon \rangle$ and $\langle \varepsilon -\hbar \omega 
|f|\varepsilon \rangle$. Since both the density of states and 
the probability $\rho_D (\varepsilon)$ are non-negative, the 
scalar product defined by eq.\ (\ref{11}) satisfies all the usual 
requirements of the scalar product. Using the standard notation 
for this product, we can write (\ref{11}) simply as 
\begin{equation} 
S_{fq}(\omega) = <f|q> \, .
\label{12}  \end{equation} 

Spectral densities of the back-action and the output noise can be 
also expressed in terms of this scalar product. The total spectral 
densities $S_q$, $S_f$ are defined as 
\begin{equation}  
S_q(\omega) =(S_{qq}(\omega) +S_{qq}(-\omega))/2 , \, \;\;\; 
S_{qq}(\omega) =\frac{1}{2\pi} \int d\tau e^{i\omega \tau} 
\langle qq(\tau) \rangle_D \, , 
\label{10}  \end{equation} 
with similar equations for $S_f(\omega)$. Writing the correlators 
$S_{qq}(\omega)$ and $S_{ff}(\omega)$ in the same way as 
$S_{fq}(\omega)$ (\ref{11}) in the basis of the detector energy 
eigenstates one sees immediately that they again can be expressed 
very simply in the language of the introduced scalar product: 
\begin{equation}
S_{qq}(\omega) = <q|q>\, , \;\;\;  S_{ff}(\omega) = <f|f>\, .
\label{13}  \end{equation}
Schwarz inequality for this scalar product gives, when applied to 
eqs.\ (\ref{12}) and (\ref{13}): 
\begin{equation}
S_{ff}(\omega)S_{qq}(\omega) \geq \mid S_{fq}(\omega) \mid^2 \, .
\label{14}  \end{equation}

Finally, combining eq.\ (\ref{9}), inequality (\ref{14}), similar 
inequality for components of the correlators at frequency $-\omega$, 
and our assumption that the noise spectral densities $S_q(\omega) $ 
and $S_f(\omega) $ are constant at low frequencies, we see that
\[ \lambda = - 4\pi \mbox{Im} S_{fq} /\hbar \, , \] 
and therefore 
\begin{equation}
|\lambda| \leq \frac{4\pi}{\hbar} [S_fS_q - (\mbox{Re} S_{fq})^2 
]^{1/2} \, , 
\label{15}  \end{equation}
where all spectral densities and the response coefficient are now taken 
at low frequencies. This inequality is the main result of application 
of the linear response theory to the measurement problem. 

One consequence of the inequality (\ref{15}) is the proof of the 
relation (\ref{8}) between the back-action dephasing rate and the 
measurement time. It follows directly from eq.\ (\ref{15}) that 
when the detector is ``symmetric'', i.e. $\mbox{Re} S_{fq} =0$, so that 
there are no classical correlations between the back-action noise and 
the output noise, inequality (\ref{15}) transforms eq.\ (\ref{7}) 
directly into the (\ref{8}). (The case of an asymmetric detector is 
discussed below.) In this derivation of (\ref{8}), this  inequality 
appears to be the consequence of some relation between the 
correlators appearing in the linear response theory. From the 
perspective of the measurement problem it can be interpreted in 
a much broader sense. Inequality (\ref{8}) shows that in general 
the dephasing of the measured system by a detector can be arbitrary 
strong without providing any information on the system. However, 
acquisition of information about the state of the system creates 
some minimum dephasing that suppresses coherence between the 
eigenstates of the measured observable. The fact that such a 
minimum exists is dictated by the basic principle of quantum 
mechanics which requires the successful measurement of an 
observable to localize the system in one of the eigenstates of this 
observable. If the detector is such that it is causing only this 
minimum back-action dephasing dictated by quantum mechanics, it is 
typically referred to as ``quantum-limited'' or ``ideal'' detector.   

Interplay between information acquisition by the detector  
and back-action dephasing of the measured system has a simple  
form of inequality (\ref{8}) only in the situation of 
measurement of a static system. When the system Hamiltonian 
$H_S$ is non-vanishing, this interplay is affected by the dynamics 
of the system and in general manifests itself less directly. One 
studied example of this \cite{b11,b12} is the measurement of 
a {\em two-state system}. Coherent quantum oscillations in this system 
are transformed by measurement into classical oscillations of the 
detector output. In this case, the trade-off between the back-action 
dephasing and information acquisition is reflected in the height of 
the oscillations peak $S_m$ in the output spectrum relative to the 
output noise $S_q$. Inequality (\ref{8}) limits then the peak 
height: $S_m/S_q\leq 4$, with eaquality reached by measurement with 
a symmetric quantum limited detector. 

Another system which has been studied in this context is {\em 
harmonic oscillator} \cite{b14,b6}. Similarly to the two-state 
system, dynamics 
of the measurement process is reflected in this case in the frequency 
spectrum of the detector output, which contains the detector output 
noise and a peak at the oscillator frequency. The trade-off between 
the detector back-action and information gain becomes relevant if one 
asks a question what is the minimum contribution of the detector 
noise to the spectrum. Qualitatively, there are two ways in which the 
detector noise contributes to the spectrum. The detector output noise 
gives direct contribution to the spectrum, while the back-action noise 
contributes to the output spectrum indirectly, by inducing additional 
oscillations at the detector input that are transformed to the output 
by the detector response. Minimization of the total noise contribution 
(reduced to the detector input and normalized to the zero-point 
spectrum of the oscillator in units of $\hbar/2$) shows that 
the minimum total noise is given by the ``energy sensitivity'' 
$\epsilon$:
\begin{equation}
\epsilon = \frac{2\pi }{|\lambda|} [S_f S_q -(\mbox{Re}S_{qf})^2]^{1/2} 
\, . 
\label{16}  \end{equation}
Energy sensitivity has the dimension of action, and relation 
(\ref{15}) between the parameters of the linear response theory 
translates into the following limitation on $\epsilon$:
\begin{equation} 
\epsilon \geq \hbar/2 \, . 
\label{17}  \end{equation}

Inequality (\ref{17}) can be expressed qualitatively as a 
statement that the detector, when measuring a harmonic oscillator, 
adds at least half an excitation quantum of this oscillator to the 
measured signal. This inequality was obtained for the first time in a 
different physical context of linear amplification of electromagnetic 
radiation, where this interpretation has direct meaning -- see, e.g., 
the papers on quantum noise in linear amplifies reprinted in 
\cite{b1}, also \cite{b13} and references therein. The magnitude of 
the signal 
in this context is described naturally by the number of photons in a 
mode and amplification is characterized by the dimensionless photon 
gain: the ration of these numbers in the output and input modes. 
Relation between such a linear amplifier and the linear detector 
considered in this work is established by the fact that in the limit 
of large gain, the input 
signal of the amplifier which in general can be quantum (i.e., contain 
only few photons) is transformed into the output signal which contains 
number of photons much larger than one and is in this sense 
classical. This means that the amplifier in this regime 
acts essentially as a detector, since after amplification of the 
signal to the classical level, it can be dealt with without 
any restrictions. 

The last point to be made here concerns the case of an asymmetric 
detector. Such a detector is defined by the nonvanishing classical 
correlations between the output and back-action noises described 
by the real part of the spectral density $S_{fq}$. Comparison of 
eq.\ (\ref{7}) for  
the trade-off between the measurement time and the back-action 
dephasing, and eq.\ (\ref{17}) for the energy sensitivity shows 
that they behave differently in the case of non-vanishing 
$\mbox{Re}S_{fq}$. While the energy sensitivity (\ref{17}) 
can reach its quantum limit even in asymmetric detectors with 
$\mbox{Re}S_{fq} \neq 0$, the $\tau_m \Gamma_d$ product (\ref{7}) 
in this situation can only be larger than its quantum limit $1/2$. 
The origin of this discrepancy is that in asymmetric detectors 
the output noise contains some information about the back-action 
noise which should be utilized to reach the optimum 
quantum-limited performance of the detector. In the simple 
treatment that lead to inequality (\ref{8}), this information 
is lost, and as a result the dephasing is increased beyond the minimum 
required by the quantum mechanics. In the case of measurement of 
the harmonic oscillator, the noise minimization procedure 
implicitly makes use of this information, since the precise 
frequency where the noise is minimum depends on the 
correlation strength $\mbox{Re}S_{fq}$. In the case of 
information/dephasing trade-off, the information provided by the 
correlations contained in non-vanishing $\mbox{Re}S_{fq}$ can 
also be used to approach the quantum limit of dephasing. 
Conceptually (and probably also practically) the simplest way 
to use the information contained in the output noise of the 
detector to reduce the back-action dephasing is to apply the 
output noise with an appropriate transfer coefficient to the 
measured system together with the back-action noise. Although 
physically this procedure might require a rather complicated 
set-up that actually applies part of the detector output to the 
measured system, formally this is achieved by simply redefining 
the back-action force of the detector: 
\[ f \rightarrow f-(\mbox{Re}S_{fq}/S_q) o \equiv f' \, . \]
It is straightforward to see that the rate $\Gamma_d'$ of back-action 
dephasing created by the redefined force $f'$ in measurement of 
a static system can indeed be equal to the quantum minimum of 
$1/2\tau_m$ even for asymmetric detectors with $\mbox{Re}S_{fq} 
\neq 0$, if the relation (\ref{15}) between the linear response 
properties of this detector is satisfied as equality. The 
same conclusion can be reached for measurement of the two-state 
systems \cite{b15}. This provides one more example of the fact that 
many physical characteristics of the measurement process can be 
understood conveniently as dynamics of information. 

\vspace*{3ex} 

\hspace*{-1.7em} {\bf Acknowledgements} 

\hspace*{-1.7em} The author would like to thank E.V. Sukhorukov for 
useful discussion of the results and for drawing his attention to Ref.\ 
\cite{b34}. This work was supported in part by AFOSR and ARDA.

\end{document}